# Medical device surveillance with electronic health records

Alison Callahan[1†], Jason A Fries[1,2†], Christopher Ré[2], James I Huddleston III[3]; Nicholas J Giori[3,4], Scott Delp[5], Nigam H Shah[1]

[1]Center for Biomedical Informatics Research, Stanford University, 1265 Welch Road, Stanford, California, USA 94305

[2]Department of Computer Science, Stanford University, 353 Serra Mall, Stanford, California, USA 94305

[3]Department of Orthopaedic Surgery, School of Medicine, Stanford University, 450 Broadway Street, Redwood City, California, USA 94063

[4]Veterans Affairs Palo Alto Health Care System, 3801 Miranda Avenue, Palo Alto, California, USA 94304

[5]Department of Bioengineering, Stanford University, 318 Campus Drive, Stanford, California USA 94305

†These authors contributed equally to this work.

Correspondence to: Alison Callahan acallaha@stanford.edu.

# Abstract

Post-market medical device surveillance is a challenge facing manufacturers, regulatory agencies, and health care providers. Electronic health records are valuable sources of real world evidence for assessing device safety and tracking device-related patient outcomes over time. However, distilling this evidence remains challenging, as information is fractured across clinical notes and structured records. Modern machine learning methods for machine reading promise to unlock increasingly complex information from text, but face barriers due to their reliance on large and expensive hand-labeled training sets. To address these challenges, we developed and validated state-of-the-art deep learning methods that identify patient outcomes from clinical notes without requiring hand-labeled training data. Using hip replacements —one of the most common implantable devices— as a test case, our methods accurately extracted implant details and reports of complications and pain from electronic health records with up to 96.3% precision, 98.5% recall, and 97.4% F1, improved classification performance by 12.7-53.0% over rule-based methods, and detected over 6 times as many complication events compared to using structured data alone. Using these additional events to assess complication-free survivorship of different implant systems, we found significant variation between implants, including for risk of revision surgery, which could not be detected using coded data alone. Patients with revision surgeries had more hip pain mentions in the post-hip replacement, pre-revision period compared to patients with no evidence of revision surgery (mean hip pain mentions 4.97 vs. 3.23; t = 5.14; p < 0.001). Some implant models were associated with higher or lower rates of hip pain mentions. Our methods complement existing surveillance mechanisms by requiring orders of magnitude less hand-labeled training data, offering a scalable solution for national medical device surveillance using electronic health records.

# Introduction

For individuals implanted with the most widely used medical devices today, including pacemakers, joint replacements, breast implants, and more modern devices, such as insulin pumps and spinal cord stimulators, effective pre-market assessment and post-market surveillance are neccessary[1] to assess their implants' safety and efficacy. Device surveillance in the United States relies primarily on spontaneous reporting systems as a means to document adverse events reported by patients and providers, as well as mandatory reporting from manufacturers and physicians to the Food and Drug Administration (FDA). As a result, device-related adverse events are significantly underreported — by some estimates as little as 0.5% of adverse event reports received by the FDA concern medical devices[2]. Recent adoption of universal device identifiers[3,4] will make it easier to capture device details, and to report events using these identifiers, but linking devices to patient outcomes remains a challenge. Unique device identifiers have not been used historically (longitudinal data are essential to monitor device safety over time) and many relevant patient outcomes are simply not recorded in a structured form amenable for analysis. The International Consortium of Investigative Journalists recently analyzed practices of medical device regulation around the world and dissemination of device recalls to the public, sharing their findings in a report dubbed the 'Implant Files'[5]. One issue highlighted in their report is the significant challenge of tracking device performance and reports of adverse events.

Observational patient data captured in electronic health records (EHRs) are increasingly used as a source of real world evidence to guide clinical decision making, ascertain patient outcomes, and assess care efficacy[6–8]. Clinical notes in particular are a valuable source of real world evidence because they capture many complexities of patient encounters and outcomes that are underreported or absent in billing codes. Prior work has demonstrated the value of EHRs, and clinical notes specifically, in increasing the accuracy[9] and lead time[10] of signal detection for post-market drug safety surveillance. The hypothesis motivating this work is that *EHRs, and evidence extracted from clinical notes in particular, can enable device surveillance and complement existing sources of evidence used to evaluate device performance*.

EHR structure and content varies across institutions, and technical challenges in using natural language processing to mine text data[11] create significant barriers to their secondary use. Pattern and rule-based approaches are brittle and fail to capture many complex medical concepts. Deep learning methods for machine reading in clinical notes[12–15] outperform these approaches to substantially increase event detection rates, but require large, hand-labeled training sets[16,17]. Thus recent research has led to new methods that use *weak supervision*, in the form of heuristics encoding domain insights, to generate large amounts of imperfectly labeled training data. These weakly supervised methods can exceed the performance of traditional supervised machine learning while eliminating the cost of obtaining large quantities of hand-labeled data for training[18,19]. Unlike labeled clinical data, these heuristics, implemented as programmatic *labeling functions*, can be easily modified and shared across institutions, thus overcoming a major barrier to learning from large quantities of healthcare data while protecting patient privacy.

In this work, we demonstrate the use of such novel methods to identify real-world patient outcomes for the one of the most commonly implanted medical devices: the hip joint implant. More than one million joint replacements are performed every year in the United States and rates are increasing, including in patients under 65 years of age[20]. Primary joint replacements are expensive[21], and implant failure incurs significant financial and personal cost for patients and the health care system. Recent recalls for metal-on-metal hip replacements[22] highlight the need for scalable, automated methods for implant surveillance[23]. While major adverse events such as deaths or revision surgery are typically reported, comprehensive device surveillance also needs to track outcomes that are reported less often, such as implant-related infection, loosening, and pain. Pain is a primary indication for undergoing a hip replacement, yet studies have found that more than 25% of individuals continue to experience chronic hip pain after replacement[24,25]. Pain is also an early indicator of the need for revision surgery[26], and thus a key implant safety signal.

Motivated by the potential impact and scale of this surveillance problem, we applied weak supervision and deep learning methods using the Snorkel framework[19], to identify reports of implant-related complications and pain from clinical notes. After validating these machine reading methods, we combined data from clinical notes with structured EHR data to characterize hip implant performance in the real world, demonstrating the utility of using such text-derived evidence. We show that our methods substantially increase the number of identified reports of implant-related complications, that our findings complement previous post-approval studies, and that this approach could improve existing approaches to identify poorly performing implants.

# Methods

We developed machine reading methods to analyze clinical notes and identify the implanted device used for a patient's hip replacement, as well as identify mentions of implant-related complications and patient-reported pain. We evaluated these methods' ability to (1) accurately identify implant-related events with a minimum of hand-labeled training data; and (2) map identified implants to unique identifiers -- a process called *canonicalization* -- to automatically replicate the contents of a manually curated joint implant registry.

We then combined the data produced by our machine learning methods with structured data from EHRs to: (1) compare complication-free survivorship of implant models and (2) derive insights about associations between complications, pain, future revision surgery and choice of implant system.

In the following sections, we first describe how we identified the patient cohort analyzed in our study. When then describe in detail our novel weakly supervised machine reading methods. Lastly, we describe our analysis of the performance of implant systems based on real world evidence produced by these methods, using Cox proportional hazards models, negative binomial models, and the t-test.

Our study was approved by the Stanford University Institutional Review Board with waiver of informed consent, and carried out in accordance with HIPAA guidelines to protect patient data.

## Cohort construction

We queried EHRs of roughly 1.7 million adult patients treated at Stanford Health Care between 1995 and 2014 to identify patients who underwent primary hip replacement and/or revision surgery. To find records of primary hip replacement surgery, we used ICD9 procedure code 81.51 (total hip replacement) and CPT codes 27130 (total hip arthroplasty) and 27132

(conversion of previous hip surgery to total hip arthroplasty). To find structured records of hip replacement revision surgery, we used ICD9 procedure codes 81.53 (Revision of hip replacement, not otherwise specified), 00.70, 00.71, 00.72 and 00.73 (Revision of hip replacement, components specified), and CPT codes 27134 (Revision of total hip arthroplasty; both components, with or without autograft or allograft), 27137 (Revision of total hip arthroplasty; acetabular component only, with or without autograft or allograft) and 27138 (Revision of total hip arthroplasty; femoral component only, with or without allograft).

We identified 6,583 patients with records of hip replacement surgery, of which, 386 (5.8%) had a coded record of at least one revision after the primary surgery. For all patients in the resulting cohort, we retrieved the entirety of their structured record (procedure and diagnosis codes, medication records, vitals etc.), their hip replacement operative reports, and all clinical notes. We were able to retrieve operative reports for 5,801 (88%) hip replacement patients.

## Machine Reading

### Background on weak supervision

In supervised machine learning, experts employ a wide range of domain knowledge to label ground truth data. *Data programming*[18,19] is a method of capturing this labelling process using a collection of imperfect heuristics or *labeling functions* which are used to build large training sets. Labeling functions encode domain expertise and other sources of indirect information or *weak supervision* (e.g., knowledge bases, ontologies) to vote on a data item's possible label. These functions may overlap, conflict, and have unknown accuracies. Data programming unifies these noisy label sources, using a generative model to estimate and correct for the unobserved accuracy of each labeling source and assign a single probabilistic label to each unlabeled input sample. This step is unsupervised and requires no ground truth data. The generative model is used to programmatically label a large training set, to learn a discriminative model such as a neural network. By training a deep learning model, we transform rules into learned feature representations, which allows us to generalize beyond the original labeling heuristics, resulting in improved classification performance.

The probabilistic labels generated via data programming results in a large training set which is referred to as "weakly labeled". The discriminative model gets as input the original text sentences in the training set and the probabilistic labels generated by the data programming step. Because the label provided in such a supervised learning setup is "weak", the process is also called "weak supervision" and the resulting discriminative model referred to as a weakly supervised model. We describe each step of the process (Figure 1) in the following sections.

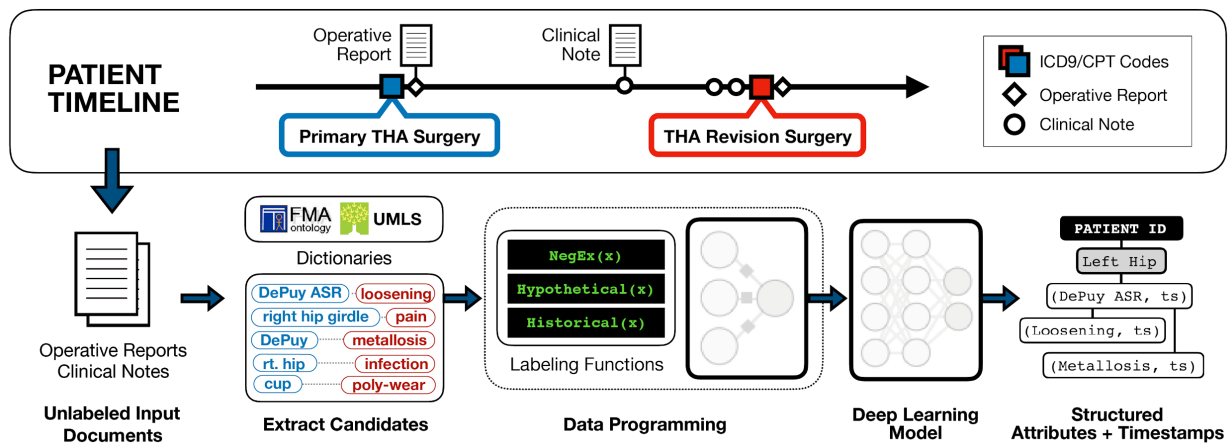

**Figure 1.** Overview of our machine reading pipeline. *Top*: Each patient's EHR is processed to extract the date of primary hip replacement surgery, any coded record of revision surgeries, and all clinical and operative notes. *Bottom*: From the patient's coded data and primary hip replacement operative report, we tagged all mentions of implants, complications, and anatomical locations. We defined pairs of relation candidates from these sets of entities, and labeled them using *data programming* via the Snorkel framework. These labeled data were then used to train a deep learning model. When applied to unseen data, the final model's final output

consists of timestamped, structured attribute data for implant systems, implant-related complications, and mentions of pain.

## Implant type, implant-related complications, and pain event definitions

We defined 3 entity/event types : (1) implant manufacturer/model entities, e.g., "Zimmer VerSys"; (2) implant-related complications, e.g., "infected left hip prosthetic"; and (3) patient-reported pain at a specific anatomical location, e.g., "left hip tenderness". The latter two, implant complications and patient-reported pain, involve multiple concepts and were formulated as *relational inference* tasks[27] where a classifier predicts links between two or more entities or *concepts* found anywhere in a sentence, i.e., `R = `$r$`(c`$_i$`...c`$_N$`)`, where $r$ is a relation type and `c`$_i$ is a concept mention. This formulation allowed our classifier learn complex semantic relationships between sentence entities, using statistical inference to capture a wide range of meaning in clinical notes. This approach enabled identifying nuanced, granular information, such as linking pain and complication events to specific anatomical locations and implant subcomponents (e.g, "*left acetabular cup* demonstrates extreme *liner wear*"). Therefore, complication outcomes were represented as entity pairs of `(complication, implant)` and pain outcomes as pairs of `(pain, anatomy)`.

We considered all present positive mentions of an event —i.e., the event was contemporaneous with a note's creation timestamp— as positive examples. Historical, negated and hypothetical mentions were labeled as 'negative'. Implant complications were further broken down into 6 disjoint subcategories: *revision, component wear, mechanical failure, particle disease, radiographic abnormality, and infection* (see supplement for examples)*.* These categories correspond to the removal of specific hardware components as well as common indicators of device failure.

## Text Preprocessing

All clinical notes were split into sentences and tokenized using the spaCy[28] framework. *Implant*, *complication*, *anatomy*, and *pain* entities were identified using a rule-based tagger built using a combination of biomedical lexical resources, e.g., the Unified Medical Language System (UMLS) [29] and manually curated dictionaries (see supplement for details). Each sentence was augmented with markup to capture document-level information such as note section headers (e.g., "Patient History") and all unambiguous date mentions (e.g., "1/1/2000") were normalized to relative time delta bins (e.g., "0-1 days", "+5 years") based on document creation time. Such markup provides document-level information to incorporate into features learned by the final classification model.

Candidate relations were defined as the Cartesian product of all entity pairs defining a relation (i.e., pain/anatomy and implant/complication) for all entities found within a sentence. Candidate events and relations were generated for all sentences in all clinical notes.

## Labeling Functions

Labeling functions leverage existing natural language tools to make use of semantic abstractions or *primitives* which make it easier for domain experts to express labeling heuristics. For example, primitives include typed named entities such as *bacterium* or *procedures*, document section structure, as well as negation, hypothetical, and historical entity modifiers provided by well-established clinical natural language processing methods such as NegEx/ConText[30,31]. These primitives provide a generalized mechanism for defining labeling functions that express intuitive domain knowledge, e.g., "Implant complications found in the 'Past Medical History' section of a note are not present positive events" (Figure 2). Primitives can be imperfect and do not need to be directly provided to the final discriminative model as features. Critically, the end deep learning discriminative model automatically learns to encode primitive information using only the original text and the training signal provided by our imperfect labels.

We developed labeling functions to identify implant-related complications and pain and its anatomical location. In total, 50 labeling functions were written, 17 shared across both tasks, with 7 task-specific functions for pain-anatomy and 25 for implant-complications. Labeling functions were written by inspecting unlabeled sentences and developing heuristics to assign individual labels (supplemental figure S2). This process was iterative, using development set data (described in the next section) to refine labeling function design.

CLINICAL NOTE

**HISTORY OF PRESENT ILLNESS**:
60 yo male with infected R hip (MRSA) s/p previous hip replacement.
LTHA November 2004 demonstrates component wear.
Acetabular cup polyethylene wear is present.

**PAST MEDICAL HISTORY**:
Hx right Zimmer Biomet hip 1/1/05 complicated by infection.

**NOTE DATE**: 07/01/2008 06:11 PM

LABELING FUNCTIONS

```
def LF1_contiguous_entities(c):
    v = len(between_words(c)) == 0
    return TRUE if v else ABSTAIN

def LF2_historical(c):
    v = has_historical_attrib(c)
    return FALSE if v else ABSTAIN

def LF3_reject_section(c):
    h1 = get_section_header(c)
    v = h1 in reject_headers
    return FALSE if v else ABSTAIN
```

| LF1 | LF2 | LF3 | LABEL | CANDIDATES |
|---|---|---|---|---|
| 1 | – | – | 1 | 60 yo male with **infected** **R hip** (MRSA) s/p previous hip replacement |
| – | 0 | 0 | 0 | Hx **right Zimmer Biomet hip** 1/1/05 complicated by **infection** |
| 1 | – | – | 1 | **Acetabular cup** **polyethylene wear** is present. |
| – | – | – | 1 | Marked lucency around the **acetabular cup** is consistent with **polyethylene wear** |

**Figure 2.** Labeling function examples. Clinical notes (*top left*) are preprocessed to generate document markup, tagging entities with parent section headers and historical, hypothetical, negated attributes. Labeling functions (*top right*) use this markup to represent domain insights as simple Python functions, e.g., a function to label mentions found in "Past Medical History" as FALSE because they are likely to be historical mentions rather than a current condition. These labeling functions vote {FALSE, ABSTAIN, TRUE} on candidate relationships (bottom) to generate a vector of noisy votes for each candidate relationship. The data programming step

described in Figure 1 uses this labeling function voting matrix to learn a generative model which assigns a single probabilistic label to each unlabeled candidate relationship.

## Manually annotated datasets

We used two sources of manually annotated data to evaluate our methods: (1) the Stanford Total Joint Registry (STJR), a curated database of joint replacement patients maintained by Stanford Health Care orthopedic surgeons which, as of March 2017, contained records for 3,714 patients and (2) clinical notes from the Stanford Health Care EHR system, manually annotated by co-authors and medical doctors expressly for the purpose of evaluating our models' performance. The clinical notes manually annotated by our team consisted of three sets:

   A.  500 operative reports annotated to identify all implant mentions.
   B.  802 clinical notes annotated to identify all complication-implant mentions.
   C.  500 clinical notes annotated to identify all pain-anatomy relation mentions.

Set A was randomly sampled from the entire hip replacement cohort's operative report corpus, using the manufacturer search query "Zimmer OR Depuy". This query captured >85% of all implant components in the STJR and accounted for >90% of all implant mentions in operative reports. Sets B and C were randomly sampled with uniform probability from the entire hip replacement cohort's clinical note corpus. Ground truth labels for implant model mentions (set A) were generated by one annotator (author AC). Ground truth labels for complication-implant mentions (set B) were generated by 5 annotators (authors AC, JF, and NHS, and two medical doctors). Ground truth labels for set C were generated by 3 annotators (authors AC and JF, and one medical doctor). We used adjudication to resolve differences between annotators.

## Classification Models

For implant manufacturer/model entity extraction (e.g., "DePuy Pinnacle", "Zimmer Longevity"), we used a rule-based tagger, as we found using dictionary-based string matching was sufficiently unambiguous to achieve high precision and recall (see supplemental methods for detail on the construction of the implant dictionary used).

We used a *Bidirectional Long Short-Term Memory* (LSTM) [32] neural network with attention[33] as the discriminative model for relational inference tasks. All LSTMs used word embeddings pre-trained using FastText[34] on 8.1B tokens (651M sentences) of clinical text. Each training instance consisted of an entity pair, its source sentence, and all corresponding text markup. All models were implemented using the Snorkel framework.

## Evaluation

We evaluated our implant extraction system using the 500 manually annotated operative reports (set A, above). We evaluated our weakly supervised pain and complication LSTMs against two baselines: (1) a traditionally supervised LSTM trained using 150 hand-labeled training documents; (2) the soft majority vote of all labeling functions for a target task. To quantify the effect of increasing training set size, we evaluated the weakly supervised LSTM neural networks trained using 150 to 50,000 weakly-labeled documents. We used training, development, and test set splits of 150/19/633 from set B for implant-complication relations and 150/250/100 documents from set C for pain-anatomy relations. All models and labeling functions were tuned on the development split and results reported are for the test split. All models were scored using precision, recall, and F1-score.

We then compared the structured output of our implant extraction method to the STJR. Given a patient, an implant component, and a timestamped hip replacement surgery, *agreement* was defined as identical entries for the component in both the STJR and our system's output; *conflict* was where the implant component(s) were recorded for a hip replacement surgery in both STJR and our data but did not match; *missingness* is where a record for a hip replacement surgery was absent in either STJR or our data. We canonicalized all implant models to the level of manufacturer/model, i.e., the same resolution used by the STJR.

## Characterizing hip implant performance in the real world

We analyzed complication-free survival of implant systems using Cox proportional hazards models, controlling for age at the time of hip replacement, gender, race, ethnicity and Charlson Comorbidity Index (CCI). CCI was categorized as none (CCI = 0), low (CCI = 1), moderate (CCI = 2) or high (CCI ≥ 3). For patients with multiple complications, we calculated survival time from primary hip replacement to first complication. We performed this analysis for a composite outcome of 'any complication', and also for each class of complication extracted by our text processing system.

To quantify association between implant systems and hip pain, we fit a negative binomial model to the frequency of hip pain mentions in the year post-hip replacement. The model included the following covariates: implant system, age at the time of hip replacement, gender, race, ethnicity, frequency of hip pain mentions in the year prior to surgery (to account for baseline levels of hip pain), and follow-up time post-THA (to a maximum of one year). For the subset of patients with body mass index (BMI) data available, we also fit models that included BMI as a covariate.

We specified implant system as an interaction term between femoral and acetabular components, grouping infrequently occurring interactions into a category "other". The frequency cutoff for inclusion in the "other" category was chosen in a data-driven manner using Akaike information criterion (AIC) to find the cutoff resulting in best model fit. We specified the most frequent system (Zimmer Trilogy acetabular + VerSys femoral component) as the reference interaction term.

To assess the association between revision and hip pain, we tested the hypothesis that hip-specific pain mentions were more frequent in patients with a coded record of revision compared to those who without. We used the two-sided t-test, controlling for post-implant follow-up time.

For all implant system analyses, we restricted the analysis to the 2,704 patients with a single hip replacement, a single femoral component and a single acetabular component.

All statistical analyses were performed using R 3.4.0.

## Code availability

Code, dictionaries, and labeling function resources required to extract implant details, complication-implant and pain-anatomy relations are publicly available at https://github.com/som-shahlab/ehr-rwe. Included are Jupyter Notebooks that run the complication-implant and pain-anatomy inference code end-to-end on MIMIC-III[35] notes.

# Results

We first describe the cohort analyzed in our study and report the performance of our implant extraction method and weakly supervised relation extraction methods in comparison to traditionally supervised models. We then compare the ability of our method to replicate the contents of the manually curated STJR joint implant registry. Lastly, we present our findings when combining the evidence produced by our extraction methods with structured data from EHRs to characterize hip implant performance in the real world, demonstrating the utility of such text processing methods.

## Cohort summary

We identified 6,583 patients with records of hip replacement surgery. The hip replacement cohort was 55.6% female, with an average age at surgery of 63 years and average follow-up time after replacement of 5.3 years (± 2.1 years). Of these 6,583 patients, 386 (5.8%) had a coded record of at least one revision surgery. For patients who had a revision surgery, the average age at primary replacement surgery was 57.9 years and average follow-up time was 10.5 years (± 3.0 years).

## Performance of machine reading methods

The performance of our implant, complication, and pain extraction methods based on test sets of manually annotated records is summarized in Table 1.

**Table 1.** Precision, recall and F1 of our best performing entity and relation extraction methods.

| Category | Task Type | Precision | Recall | F1 | Test set size |
|---|---|---|---|---|---|
| Implant Manufacturer/Model | Entity | 96.3 | 98.5 | 97.4 | 500 |
| Pain-Anatomy | Relation | 80.2 | 82.6 | 81.4 | 100 |
| Implant-Complication | Relation | 82.7 | 62.3 | 71.1 | 633 |

Comparison of weakly supervised models to baseline models

Table 2 reports the performance of our relation extraction models compared against the baselines described in the methods. Using weakly-labeled training data substantially improved performance in all settings. Weakly supervised LSTM models improved 9.2 and 24.7 F1 points over the soft majority vote baseline, with 17.8 and 29.9 point gains in recall. Directly training an LSTM model with just the 150 hand-labeled documents in the training set increased recall (+13.6 and +15.4 points) over the majority vote for both tasks, but lost -8.9 and -30.9 points in precision. These results show that training using large amount of generated, imperfectly labeled data is beneficial in terms of recall, while preserving precision.

For subcategories of implant complications, F1 was improved by 16.2 to 496% for different categories except *radiographic abnormalities*, where a rule-based approach performed 18.5% better. In general, models trained on imperfectly labeled data provided substantial increases in recall, especially in cases such as *particle disease*, where rules alone had very low recall.

**Table 2.** Performance of weakly supervised relation extraction compared with baselines. Blue highlights show highest value for a given metric. Gray rows are subcategories of implant complications.

| Complication Types | Mentions | BASELINE Soft Majority Vote | | | 150 Hand-labeled documents in the training set | | | Weakly Supervised Model | | | +/- F1 |
|---|---|---|---|---|---|---|---|---|---|---|---|
| | Count | P | R | F1 | P | R | F1 | P | R | F1 | % |
| **Pain-Anatomy** | 236 | **81.4** | 64.8 | 72.2 | 72.5 | 78.4 | 75.4 | 80.2 | **82.6** | **81.4** | +12.7 |
| **Implant-Complication** | 276 | 81.7 | 32.4 | 46.4 | 50.8 | 47.8 | 49.3 | **82.7** | **62.3** | **71.1** | +53.2 |
| Revision | 63 | 74.4 | 46.0 | 56.9 | 41.7 | 68.3 | 51.8 | **75.5** | **58.7** | **66.1** | +16.2 |
| Component Wear | 48 | 71.4 | 41.7 | 52.6 | **78.9** | 31.9 | 45.5 | 72.9 | **72.9** | **72.9** | +38.6 |
| Mechanical Failure | 25 | 87.5 | 28.0 | 42.4 | 21.9 | 28.0 | 24.6 | **91.7** | **44.0** | **59.5** | +40.3 |
| Particle Disease | 65 | 80.0 | 6.2 | 11.4 | 53.8 | 32.3 | 40.4 | **97.1** | **52.3** | **68.0** | +496.5 |
| Radiographic Abnormality | 17 | **100.0** | **37.5** | **54.5** | 37.0 | 55.6 | 44.4 | 60.0 | 35.3 | 44.4 | -18.5 |
| Infection | 58 | **100.0** | 39.7 | 56.8 | 90.0 | 62.1 | 73.5 | 90.7 | **84.5** | **87.5** | +54.0 |

Figure 3 shows the impact of training set size on the weakly supervised LSTM's performance for identifying implant complications. With training set sizes greater than 10,000 weakly labeled documents, LSTM models outperformed rules across all possible classification thresholds, with a 51.3% improvement in area under the precision-recall curve (PR-AUC) over soft majority vote.

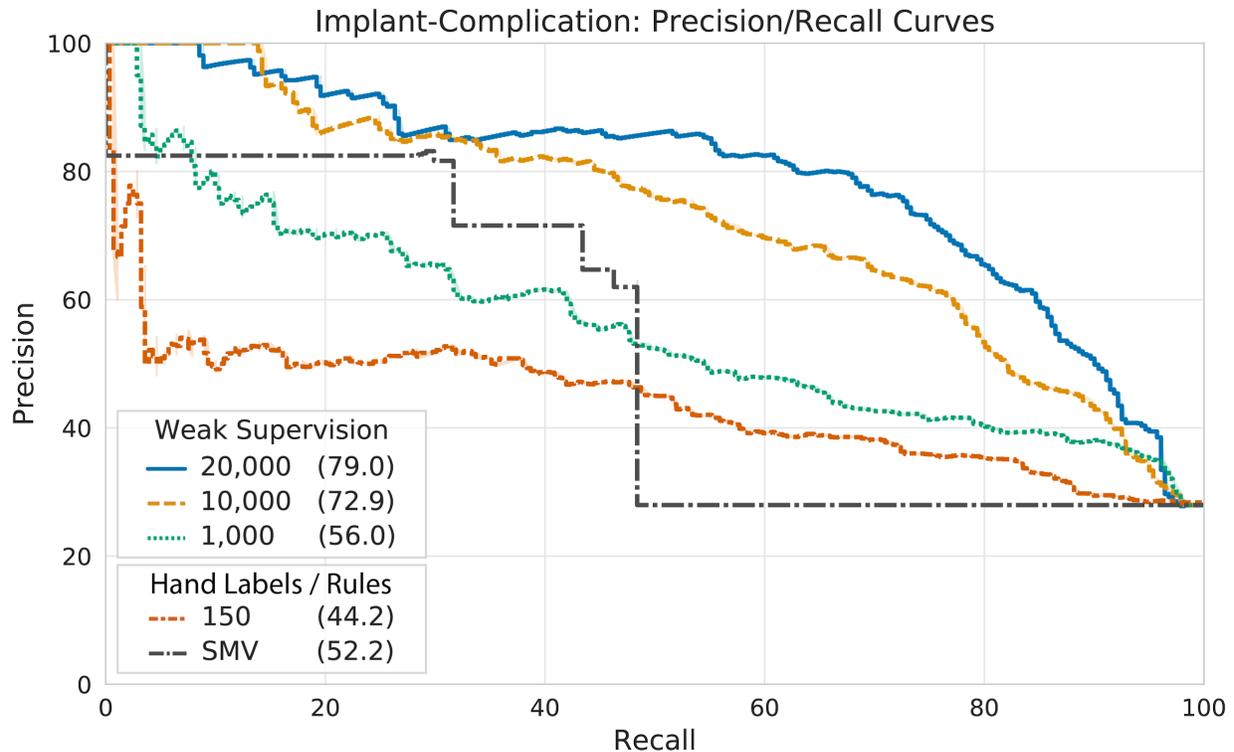

**Figure 3.** Precision Recall (PR) curves for Implant-Complication classifier performance at different training set sizes. The soft majority vote (SMV) baseline consists of a majority vote of all labeling functions applied to test data. Note how SMV favors precision at substantial cost to recall. A classifier trained directly on the 150 hand-labeled training documents failed to produce a high-performance model. When training using the generated imperfectly labeled training data, at 10k - 20k documents, the resulting supervised LSTMs are better than SMV at all threshold choices. The model trained on 20k documents performed best overall, with a 51.3% improvement over SMV in average precision (PR-AUC).

## Comparison with total joint registry

The STJR snapshot covered 3,714 patients. 2,850 STJR patients underwent surgery during the time period of this study (1995-2014). 1,877 STJR patients overlapped with our hip replacement cohort (Figure 4) and were used to evaluate model performance.

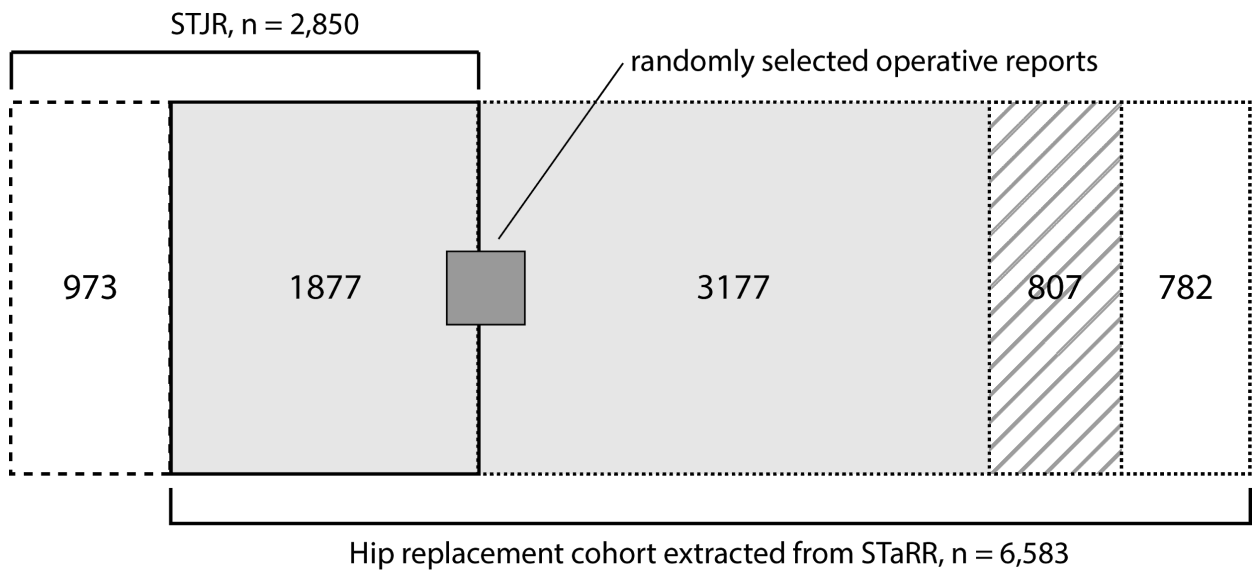

**Figure 4.** Overlap between Stanford Total Joint Registry (STJR) and the hip replacement cohort extracted from the Stanford Medicine Research Data Repository (STaRR). Of the 6,583 patients in the hip replacement cohort, 782 patients did not have an operative report (white box with dotted outline). For the 5,801 patients with operative reports, 807 did not have a mention of any implant model in their report(s) (striped box). Of the remaining 4,994 patients (light gray box), 1,877 overlapped with the STJR. Of the 477 patients whose 500 operative reports were manually annotated (dark gray box), 185 overlapped with the STJR. 973 patient records were present only in the STJR (white box with dashed outline).

For the most frequently implanted systems with ≥1 records present in the STJR (Zimmer Biomet Trilogy, DePuy Pinnacle, Depuy Duraloc, Zimmer Biomet Continuum, Zimmer Biomet RingLoc, Zimmer Biomet VerSys, Depuy AML and Zimmer Biomet M/L Taper; which corresponded to 88% of patients in the STJR/cohort overlapping set), there was 72% agreement between STJR and our system output, 17% conflict and 11% missingness.

# Hip implant system performance in the real world

In the following sections we summarize our analysis of hip implant performance using the evidence identified by our machine reading methods. First we quantify additional evidence of revision identified by our methods, and the association between implant system and risk of revision. Next, we summarize our analysis of the association between implant systems and post-implant complications generally. Lastly, we describe our findings regarding the association between implant system, hip pain, and revision. Our first two analyses used Cox proportional hazards models to measure the association between implant system and revision/post-implant complications. Our thirds analysis used a negative binomial model to investigate the association between implant system and frequency of hip pain mentions over time, and a t-test to investigate the association between hip pain and revision surgery.

## Unstructured data captures additional evidence of revision

In the analyzed subset of 2,704 patients with a single implant, only 78 had a coded record of revision surgery (as an ICD or CPT code in a billing record, for example). Considering only this evidence of revision, there was no significant association between implant system and revision-free survival (log-rank test p = 0.8; see supplementary Figure S2). When including evidence of revision extracted from clinical notes, we detected an additional 504 revision events (over 6 times as many events). In total, our methods identified 519 unique revision events (63 revision events that were detected in both coded data and text for a single patient were merged, and assigned the timestamp of whichever record occurred first). 63/78 (81%) of the coded revision events have corresponding evidence of the revision extracted from clinical notes by our system. The majority of coded records and text-derived revision events occur within 2 months of each other, suggesting good agreement between these two complementary sources of evidence when both do exist.

Analyzing the augmented revision data using a Cox proportional hazards model, different implant systems are associated with a significantly higher or lower risk of revision when controlling for age at the time of implant, race, gender, and Charlson Comorbidity Index (log-rank test p < 0.001). Figure 5 summarizes the risk of revision for implant systems when including evidence from clinical notes alongside structured records of revision.

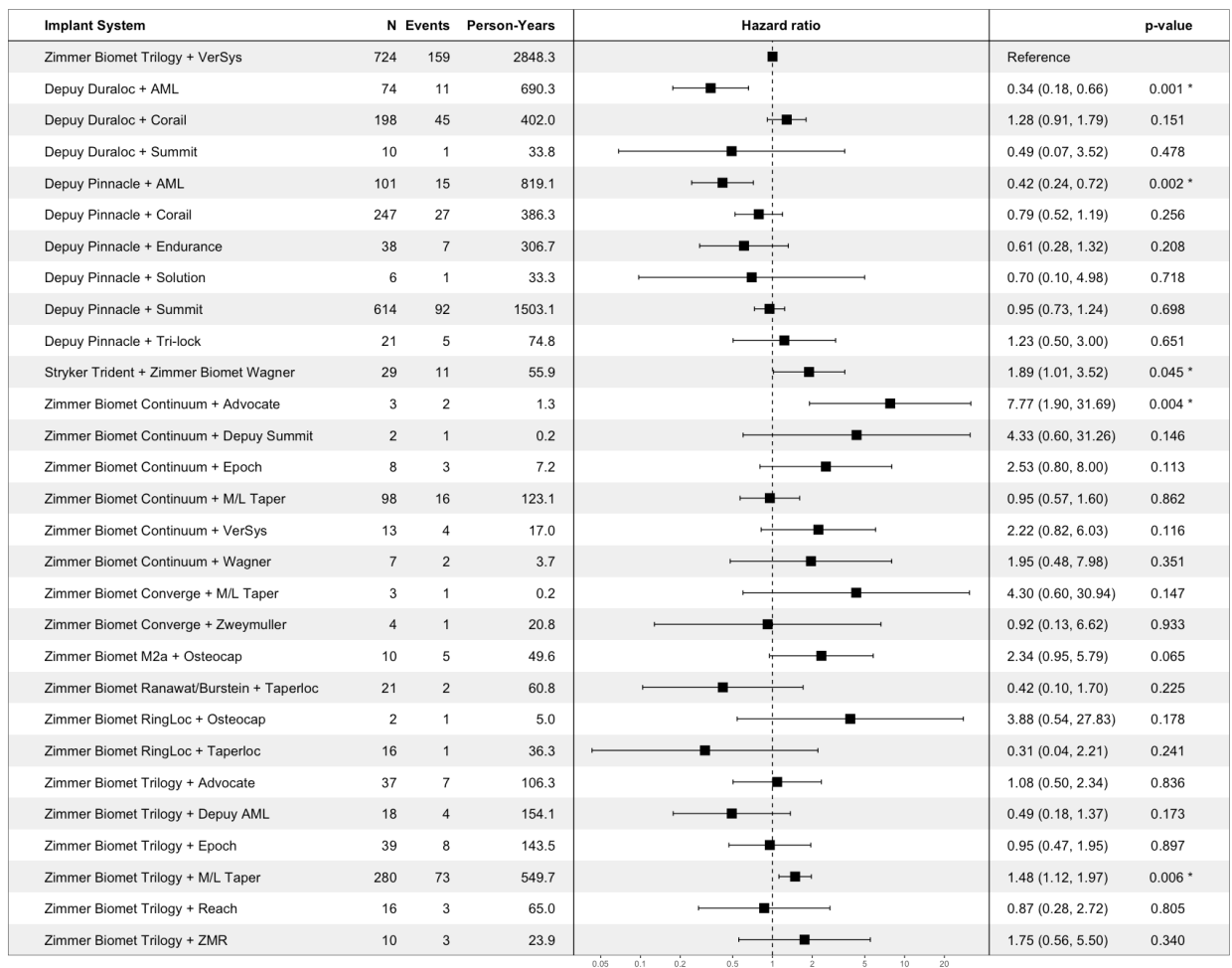

| Implant System | N | Events | Person-Years | Hazard ratio | p-value |
|---|---|---|---|---|---|
| Zimmer Biomet Trilogy + VerSys | 724 | 159 | 2848.3 | Reference | |
| Depuy Duraloc + AML | 74 | 11 | 690.3 | 0.34 (0.18, 0.66) | 0.001 * |
| Depuy Duraloc + Corail | 198 | 45 | 402.0 | 1.28 (0.91, 1.79) | 0.151 |
| Depuy Duraloc + Summit | 10 | 1 | 33.8 | 0.49 (0.07, 3.52) | 0.478 |
| Depuy Pinnacle + AML | 101 | 15 | 819.1 | 0.42 (0.24, 0.72) | 0.002 * |
| Depuy Pinnacle + Corail | 247 | 27 | 386.3 | 0.79 (0.52, 1.19) | 0.256 |
| Depuy Pinnacle + Endurance | 38 | 7 | 306.7 | 0.61 (0.28, 1.32) | 0.208 |
| Depuy Pinnacle + Solution | 6 | 1 | 33.3 | 0.70 (0.10, 4.98) | 0.718 |
| Depuy Pinnacle + Summit | 614 | 92 | 1503.1 | 0.95 (0.73, 1.24) | 0.698 |
| Depuy Pinnacle + Tri-lock | 21 | 5 | 74.8 | 1.23 (0.50, 3.00) | 0.651 |
| Stryker Trident + Zimmer Biomet Wagner | 29 | 11 | 55.9 | 1.89 (1.01, 3.52) | 0.045 * |
| Zimmer Biomet Continuum + Advocate | 3 | 2 | 1.3 | 7.77 (1.90, 31.69) | 0.004 * |
| Zimmer Biomet Continuum + Depuy Summit | 2 | 1 | 0.2 | 4.33 (0.60, 31.26) | 0.146 |
| Zimmer Biomet Continuum + Epoch | 8 | 3 | 7.2 | 2.53 (0.80, 8.00) | 0.113 |
| Zimmer Biomet Continuum + M/L Taper | 98 | 16 | 123.1 | 0.95 (0.57, 1.60) | 0.862 |
| Zimmer Biomet Continuum + VerSys | 13 | 4 | 17.0 | 2.22 (0.82, 6.03) | 0.116 |
| Zimmer Biomet Continuum + Wagner | 7 | 2 | 3.7 | 1.95 (0.48, 7.98) | 0.351 |
| Zimmer Biomet Converge + M/L Taper | 3 | 1 | 0.2 | 4.30 (0.60, 30.94) | 0.147 |
| Zimmer Biomet Converge + Zweymuller | 4 | 1 | 20.8 | 0.92 (0.13, 6.62) | 0.933 |
| Zimmer Biomet M2a + Osteocap | 10 | 5 | 49.6 | 2.34 (0.95, 5.79) | 0.065 |
| Zimmer Biomet Ranawat/Burstein + Taperloc | 21 | 2 | 60.8 | 0.42 (0.10, 1.70) | 0.225 |
| Zimmer Biomet RingLoc + Osteocap | 2 | 1 | 5.0 | 3.88 (0.54, 27.83) | 0.178 |
| Zimmer Biomet RingLoc + Taperloc | 16 | 1 | 36.3 | 0.31 (0.04, 2.21) | 0.241 |
| Zimmer Biomet Trilogy + Advocate | 37 | 7 | 106.3 | 1.08 (0.50, 2.34) | 0.836 |
| Zimmer Biomet Trilogy + Depuy AML | 18 | 4 | 154.1 | 0.49 (0.18, 1.37) | 0.173 |
| Zimmer Biomet Trilogy + Epoch | 39 | 8 | 143.5 | 0.95 (0.47, 1.95) | 0.897 |
| Zimmer Biomet Trilogy + M/L Taper | 280 | 73 | 549.7 | 1.48 (1.12, 1.97) | 0.006 * |
| Zimmer Biomet Trilogy + Reach | 16 | 3 | 65.0 | 0.87 (0.28, 2.72) | 0.805 |
| Zimmer Biomet Trilogy + ZMR | 10 | 3 | 23.9 | 1.75 (0.56, 5.50) | 0.340 |

**Figure 5. Summary of Cox proportional hazards analysis of the risk of revision for each hip implant system.** The table on the left lists the number of patients implanted with each system, the number of revision events observed for each, and the total person-years of data available. The forest plot displays the corresponding hazard ratio, with the hazard ratio (95%

confidence interval) and p-value listed in the table to the right. Note that this figure only shows implant systems for which at least one revision event was detected.

## Post replacement complications are associated with implant systems

In addition to revision, the Cox proportional hazards analysis of post-hip replacement complications indicates that some implant systems are associated with a higher or lower risk of implant-related complications overall (log-rank test $p < 0.001$) and of specific types (see supplementary results, Figures S3-S7, for specific complication types). For example, the Depuy Pinnacle (acetabular component) + AML (femoral component) system has a significantly lower risk of overall complications in comparison to the Zimmer Biomet Trilogy (acetabular) + VerSys (femoral) system, when controlling for age at the time of implant, race, gender, and Charlson Comorbidity Index (see Figure 6).

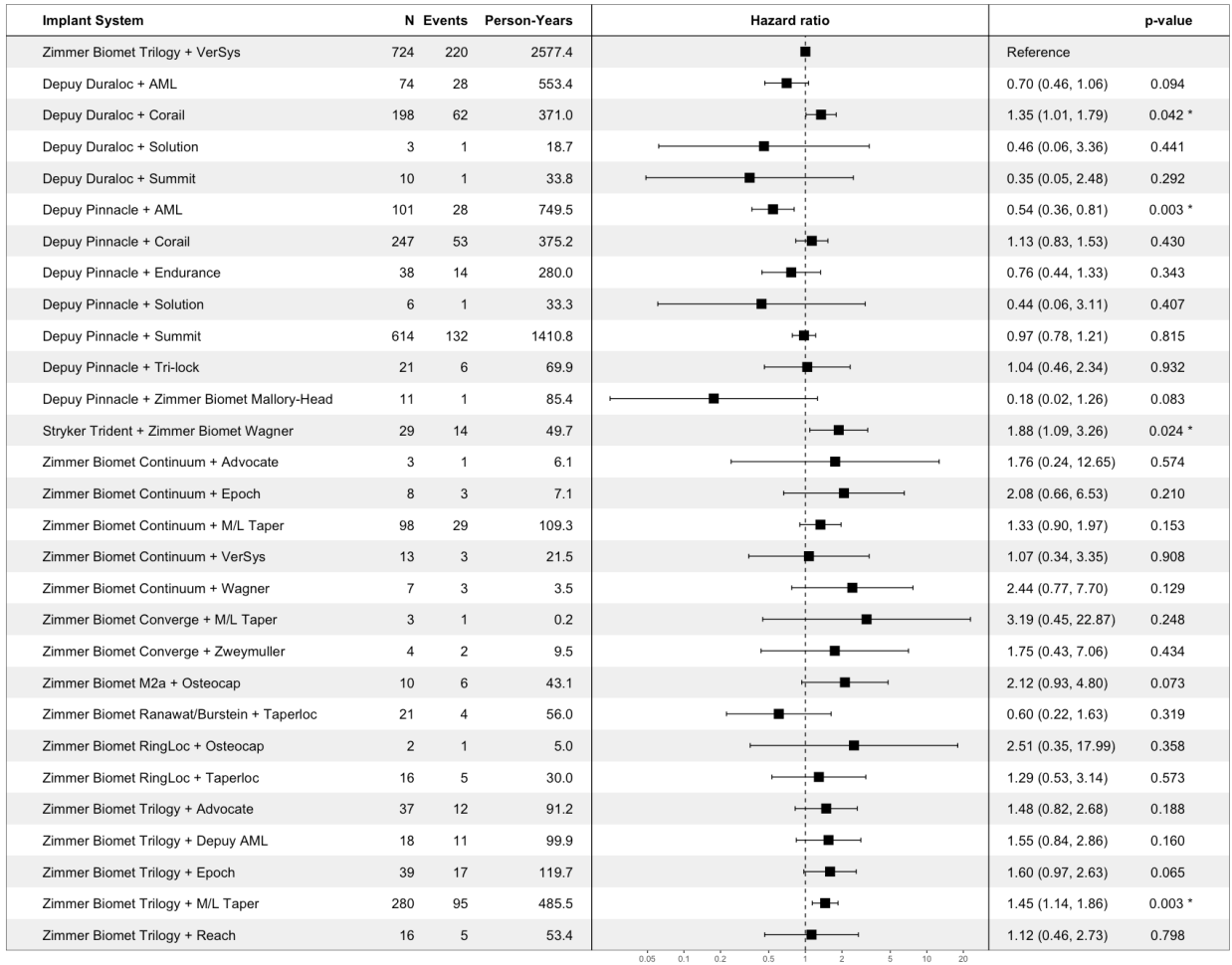

**Figure 6. Summary of Cox proportional hazards analysis of the risk of any complication for each hip implant system.** The table on the left lists the number of patients implanted with each system, the number of complication events observed for each, and the total person-years of data available. The forest plot displays the corresponding hazard ratio, with the hazard ratio (95% confidence interval) and p-value listed in the table to the right. Note that this figure only shows implant systems for which one complication event was observed.

## Post replacement hip pain is associated implant systems and revision

Our system extracted 938,222 positive pain-anatomy relation mentions from 471,985 clinical notes for 5,562 hip replacement patients, with a mean of 168 mentions per patient, and 1 mention per note. Fitting a negative binomial model to the subset of this data corresponding to the 2,704 patients with a single implant, we found that some implant systems are associated with frequency of hip pain mentions in the year following hip replacement surgery. Other covariates that also associated with frequency of post-implant hip pain are hip pain in the year prior to hip replacement, having the race category "other", unknown ethnicity, and follow-up time (see supplementary Table S1 for all model coefficients).

Table 3 lists implant systems with statistically significant negative binomial model coefficients (p ≤ 0.05), their incidence rate ratios (IRR) and 95% confidence intervals. Six systems, all manufactured by Depuy, had IRRs <1, indicating that they are associated with lower rates of hip pain mentions relative to the Zimmer Biomet Trilogy + VerSys reference system when controlling for patient demographics, pain mentions in the prior year and CCI. Five systems (4 Zimmer Biomet, 1 Depuy) have IRRs > 1, indicating that they are associated with higher rates of hip pain mentions following implant.

For 1,463 patients who had a BMI measurement within 100 days of hip replacement, we assessed whether hip pain mentions in the year following implant varies with BMI by including it in a negative binomial model alongside the above variables. We found no statistically significant association between BMI and pain in the year following hip replacement (p = 0.29).

**Table 3.** Negative binomial model-derived incidence rate ratios (IRRs), 95% confidence intervals (CI) and associated p-values for implant systems significantly associated with frequency of hip pain mentions in the year following hip replacement.

| Implant System | IRR (95% CI) | p-value |
| --- | --- | --- |
| Depuy Duraloc + AML | 0.033 (0.012 - 0.091) | < 0.001 |
| Depuy Duraloc + Corail | 1.268 (1.020 - 1.577) | 0.033 |
| Depuy Duraloc + Summit | 0.161 (0.040 - 0.652) | 0.011 |
| Depuy Pinnacle + AML | 0.097 (0.057 - 0.163) | < 0.001 |
| Depuy Pinnacle + Corail | 0.594 (0.475 - 0.742) | < 0.001 |
| Depuy Pinnacle + Endurance | 0.032 (0.010 - 0.137) | < 0.001 |
| Depuy Pinnacle + Summit | 0.369 (0.301 - 0.432) | < 0.001 |
| Zimmer Biomet Continuum + M/L Taper | 2.061 (1.561 - 2.720) | < 0.001 |
| Zimmer Biomet Ranawat/Burstein + Taperloc | 2.061 (1.174 - 3.621) | 0.012 |
| Zimmer Biomet Trilogy + M/L Taper | 1.490 (1.234 - 1.799) | < 0.001 |

Mean post-hip replacement hip pain mention frequency is significantly higher in patients who underwent revision surgery, when controlling for length of post-implant follow-up time (8.94 vs. 3.23; t = 17.60; p < 0.001). This holds true when considering only the period between primary hip replacement and revision for those patients who had revision surgery (4.97 vs. 3.23; t = 5.14; p < 0.001).

## Discussion

Post-market medical device surveillance in the United States currently relies on spontaneous reporting systems such as the FDA's Manufacturer and User Facility Device Experience (MAUDE), and increasingly on device registries[36–39]. Neither source provides a complete or accurate profile of the performance of medical devices in the real world[23]. Submitting data to registries is voluntary and not all records capture complete details on primary procedure, surgical factors, complications, comorbidities, patient reported outcomes, and radiograph[40], so existing state and health system-level registries are not comprehensive. The shortcomings of spontaneous reporting systems, which are well known in pharmacovigilance research[41,42] (including lack of timeliness, bias in reporting, and low reporting rates), also plague device surveillance efforts[43]. EHRs, with their continuous capture of data from diverse patient populations and over long periods of time, offer a valuable source of evidence for medical device surveillance in the real world, and complement these existing resources. Indeed, the FDA's new five year strategy[44] for its Sentinel post-market surveillance system prioritizes increased capture of data from EHRs. Our methods to extract evidence for device safety signal detection from EHRs directly support these efforts.

Our inference-based methods for identifying the implant manufacturer/model, pain and complications from the EHR were very accurate (F1 97.4; 81.4; 71.1) and substantially outperformed pattern and rule-based methods. Moreover, these machine learning methods can be deployed continuously, enabling near real-time automated surveillance. Comparing our method's output to a manually curated joint registry, we achieved majority agreement with existing registry records and increased coverage of hip replacement patients by 200%. Discrepancies between implant information extracted from operative reports and the corresponding registry record primarily resulted from variability in how implant system names were recorded; thus, one way our methods can augment existing registries is to standardize record capture. Our findings highlight the importance of using evidence derived from clinical notes in evaluating device performance. Our methods identified significantly more evidence of revision surgery (over 6X more events than from codes alone), augmenting what is available in structured EHR data. Our finding that text-derived evidence increases the observed rate of complications in comparison to coded data agrees with studies of the completeness and accuracy of ICD coding for a variety of diseases[45], and for comorbidities and complications of total hip arthroplasty specifically[46,47].

We found that a subset of implant systems were associated with higher or lower risk of implant-related complications in general, and of specific classes of complication. A recent meta-analysis of implant combinations[48] found no association between implant system and risk of revision, although other studies have mixed conclusions[49–54]. These studies relied on structured records of revision and registry data, and thus are complemented by our analysis of real world evidence derived from EHRs. We also found that patients with structured records of revision surgery reported more hip pain in the post replacement, pre-revision period than patients who did not. This agrees with previous findings[36,55] that pain is an early warning sign of complications that result in revision surgery.

Our approach leverages data programming to generate training sets large enough to take advantage of deep learning[32] for relational inference (which is one of the most challenging problems in natural language processing[27]). The resulting models sacrificed small amounts of precision over a rule-based approach for significant gains in recall, demonstrating their ability to capture complex semantics and generalize beyond input heuristics. This approach requires only a small collection of hand-labeled data to validate end model performance. By focusing on creating labeling functions, instead of manually labeling training data, we achieved state-of-the-art performance using code that, unlike labeled data, can be easily updated and shared across different healthcare systems.

There are several limitations to our work. We were not able to retrieve operative reports for all hip replacement patients and thus likely missed some implants. Patients had different numbers of clinical notes, at varying time points, which may not have captured their full experience of pain or complications following surgery. Our approach canonicalized implants to the level of manufacturer/model names, which only identifies broad implant families rather than the specific details provided by serial and lot numbers. However, as the use of unique device identifiers grows in popularity, our approach can incorporate this information to further differentiate the performance of implants with different design features. Our study was necessarily restricted to the implant systems used by Stanford Health Care surgeons, and thus our findings are limited to those systems. While we controlled for patient age, gender, race, ethnicity, hip pain mentions prior to surgery, and CCI in our statistical analyses, there are additional patient- and practice-specific features that may confound the complication-free survival of individual implants and rates of post-implant hip pain. These include case complexity, the orthopaedic surgeon who performed the hip replacement, their preference for specific implant systems, and their case mix. CCI is an indicator of case complexity, but joint arthroplasty-specific factors, such as indication for surgery, also contribute to patients' overall complexity. Lastly, our analysis was restricted to patients who underwent a single hip replacement, to avoid attributing a complication or pain event to the incorrect implant, in the case of patients who underwent multiple primary implant procedures.

In conclusion, we demonstrate the feasibility of a scalable, accurate, and efficient approach for medical device surveillance using EHRs. We have shown that implant manufacturer and model, implant-related complications, as well as mentions of post-implant pain can be reliably identified from clinical notes in the EHR. Leveraging recent advances in machine reading and deep learning, our methods require orders of magnitude less labeled training data and obtain state-of-the-art performance. Our findings that implant systems vary in their revision-free survival and that patients who had revision surgery had more mentions of hip pain after their primary hip replacement agree with multiple prior studies. Associations between implant systems, complications, and hip pain mentions as found in our single-site study only demonstrate the feasibility of using EHRs for device surveillance, but do not establish causality. The ability to quantify pain and complication rates automatically over a large patient population offers an advantage over surveillance systems that rely on individual reports from patients or surgeons. We believe that the algorithms described here can be readily scaled, and we make them freely available for use in analyzing electronic health records nationally.

# Acknowledgements

The authors acknowledge the contributions of Dr. Saurabh Gombar MD, Dr. Husham Sharifi MD, and Dr. James Tooley MD, who annotated clinical notes for mentions of implant-related complications, pain and anatomical location to enable evaluation of our text processing methods, and Dr. Kenneth Jung who provided guidance in carrying out statistical analyses.

# Supplemental Material

## Supplemental Methods

### Dictionary-based preprocessing

For hip implant systems, we built a dictionary of system names by querying the FDA Global Unique Device Identifier Database [56], which captures >900 hip implant components including femoral stems, femoral heads, acetabular components, and liners. For pain, we built a dictionary of 31 terms (e.g. 'pain', 'tender',) through manual inspection of notes. A complications dictionary of 452 terms was built via manual inspection of notes by clinical experts. Dictionaries were automatically expanded using an open source corpus processor [57] to capture synonyms and misspellings. The dictionary of anatomical entities consisted of all strings in the Foundational Model of Anatomy (FMA) [58], a small dictionary of informal abbreviations ("abd" -> "abdomen"), and regular expressions for standard anatomical terms of position (e.g., "*lateral* left knee").

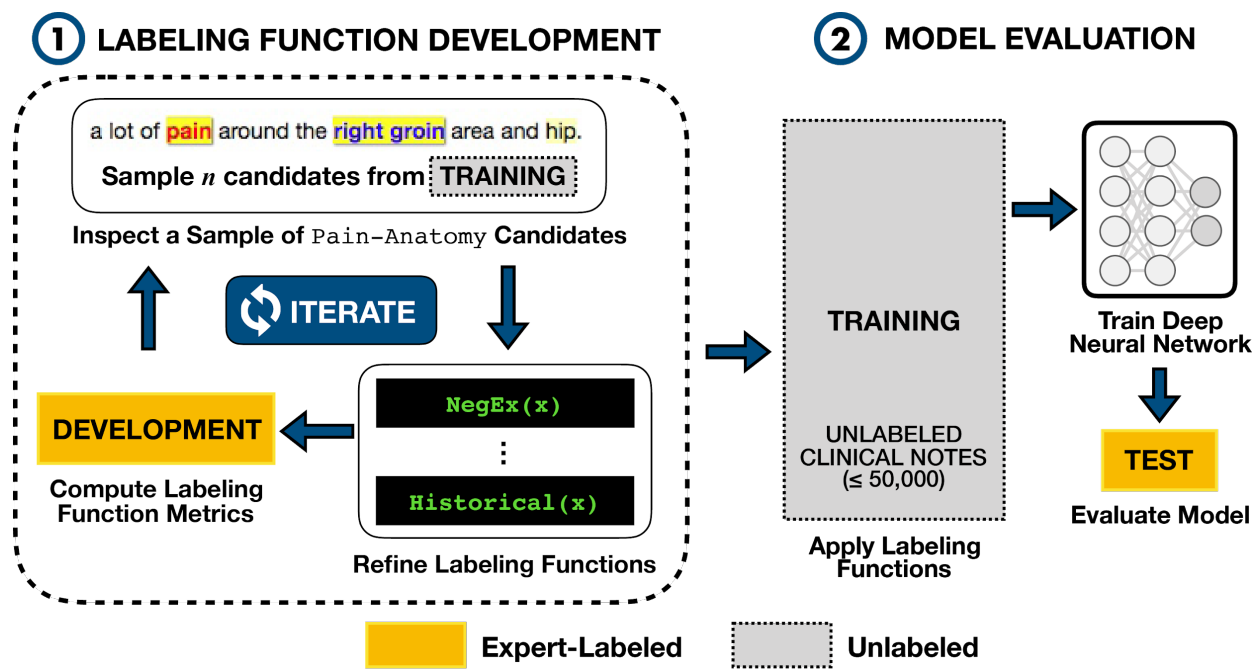

**Figure S1**. The labeling function development and model evaluation workflow. In (1) domain experts examine unlabeled candidate relationships to gain insight into writing and refining labeling functions. These functions are then empirically evaluated for accuracy, precision, recall, and F1 score on an expert-labeled development set. This is an iterative process until the desired labeling function performance is achieved on the development set. In (2) the final labeling functions are applied to a large collection of unlabeled data to generate probabilistic labels for training a deep learning model. The resulting trained model is evaluated on expert-labeled unseen test set. This approach requires orders of magnitude less hand-labeled data than what would be needed for directly training deep learning model in (2), because hand-labeled data is only used to develop labeling functions and to evaluate final model performance.

## Concept extraction models

By restricting our implant candidate extraction to the specific operative notes for each patient's THA procedure, we sufficiently disambiguated implant mentions to achieve high performance using dictionary-based string matching. Thus our implant candidates were used directly as our final implant outputs.

For pain extraction, we learned a generative model from labeling functions applied to unlabeled patient notes to create a probabilistically labeled training set. We then used this data to train a state-of-the-art *Bidirectional Long Short-Term Memory* (LSTM) [32] neural network with attention as our end discriminative model. Hyperparameter tuning was done using random search over 10 models, using a parameter grid derived from the literature (batch_size: {32, 128, 256}, dropout: {0.0, 0.25, 0.5}, emb_dim: {100, 300, 500}, output_layer_size: {50, 100, 400}, lstm_layers: {1,2,4}, learning_rate: [1e-4, 1e-2]).

For the final predicted pain events, all anatomical entities were normalized to UMLS concept unique identifiers (CUIs) using rule-based linking to the FMA. CUIs were linked to the most specific (i.e., longest distance to root node) concept in the FMA.

## Supplemental Results

### Structured revision record-free survival among implant systems

Figure S2 summarizes the risk of revision for implant systems when including evidence from structured records of revision only. Based on this data, no implant system is associated with a significantly higher or lower risk of revision. Figures S3-S7 summarize the risk of component wear, mechanical failure, particle disease, radiographic abnormality and infection (the complication subclasses detected by our extraction pipeline) for implant systems.

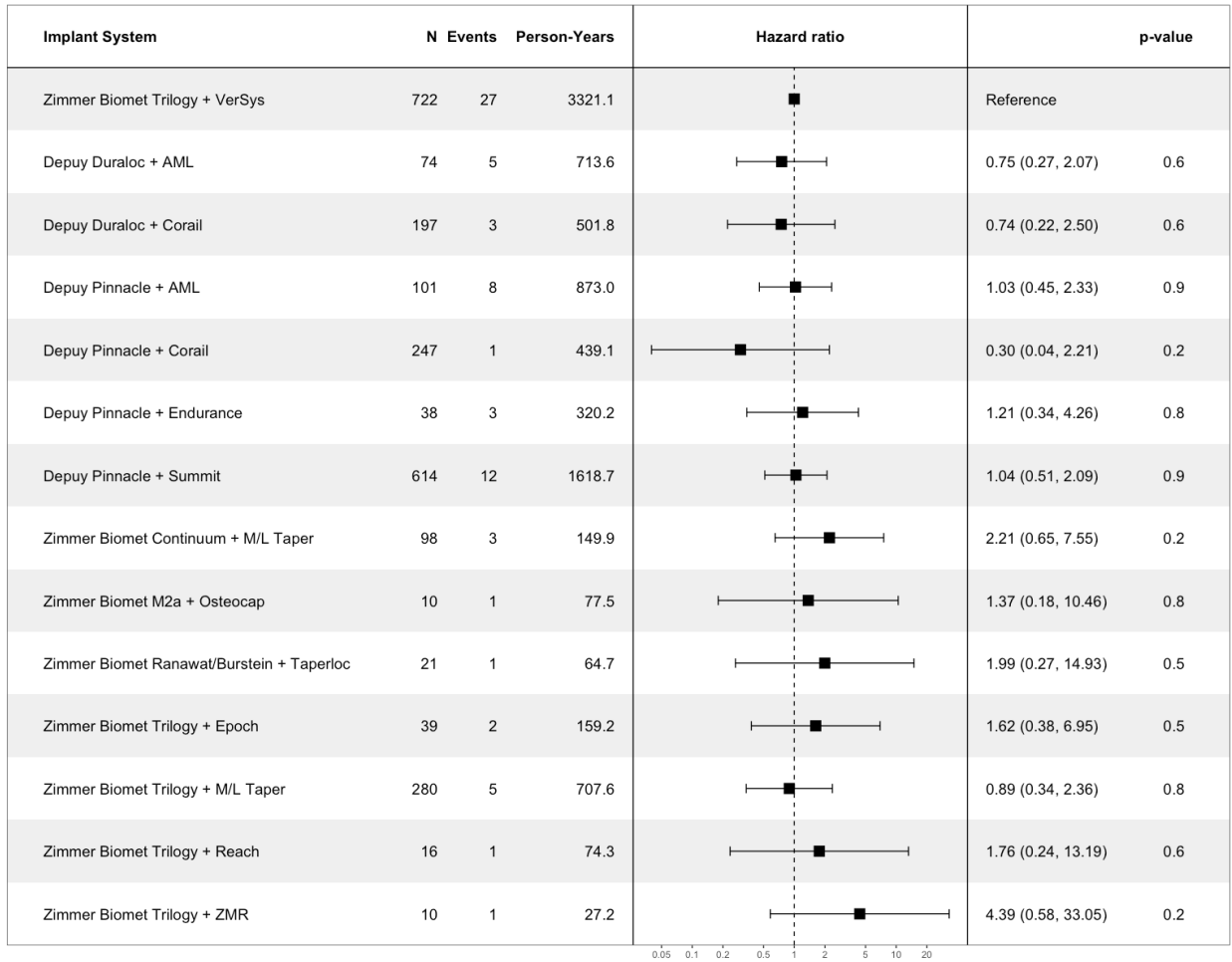

**Figure S2. Summary of Cox proportional hazards analysis of the risk of revision for each hip implant system, when including only structured records of revision.** The table on the left lists the number of patients implanted with each system, the number of revision events observed for each based on structured records only, and the total person-years of data available. The forest plot displays the corresponding hazard ratio, with the hazard ratio (95% confidence interval) and p-value listed in the table to the right. Note that this figure only shows implant systems for which at least one revision event was detected.

## Post-implant complication-free survival among implant systems

Figures S3-S7 summarize the risk of each class of post-implant complication for different implant systems, as derived by a Cox proportional hazards analysis.

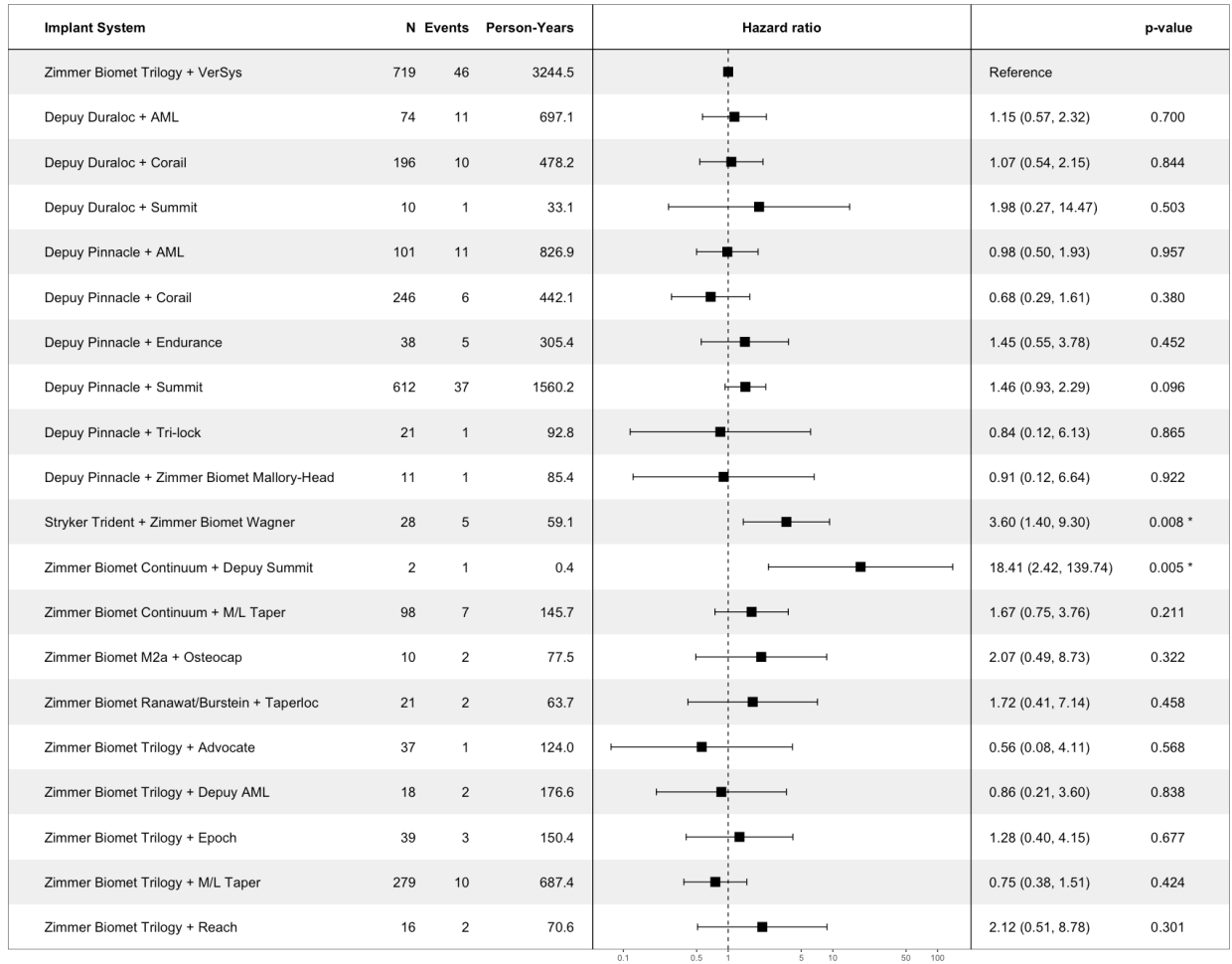

**Figure S3. Summary of Cox proportional hazards analysis of the risk of component wear for each hip implant system.** The table on the left lists the number of patients implanted with each system, the number of component wear events observed for each, and the total person-years of data available. The forest plot displays the corresponding hazard ratio, with the hazard ratio (95% confidence interval) and p-value listed in the table to the right. Note that this figure only shows implant systems for which at least one component wear event was detected.

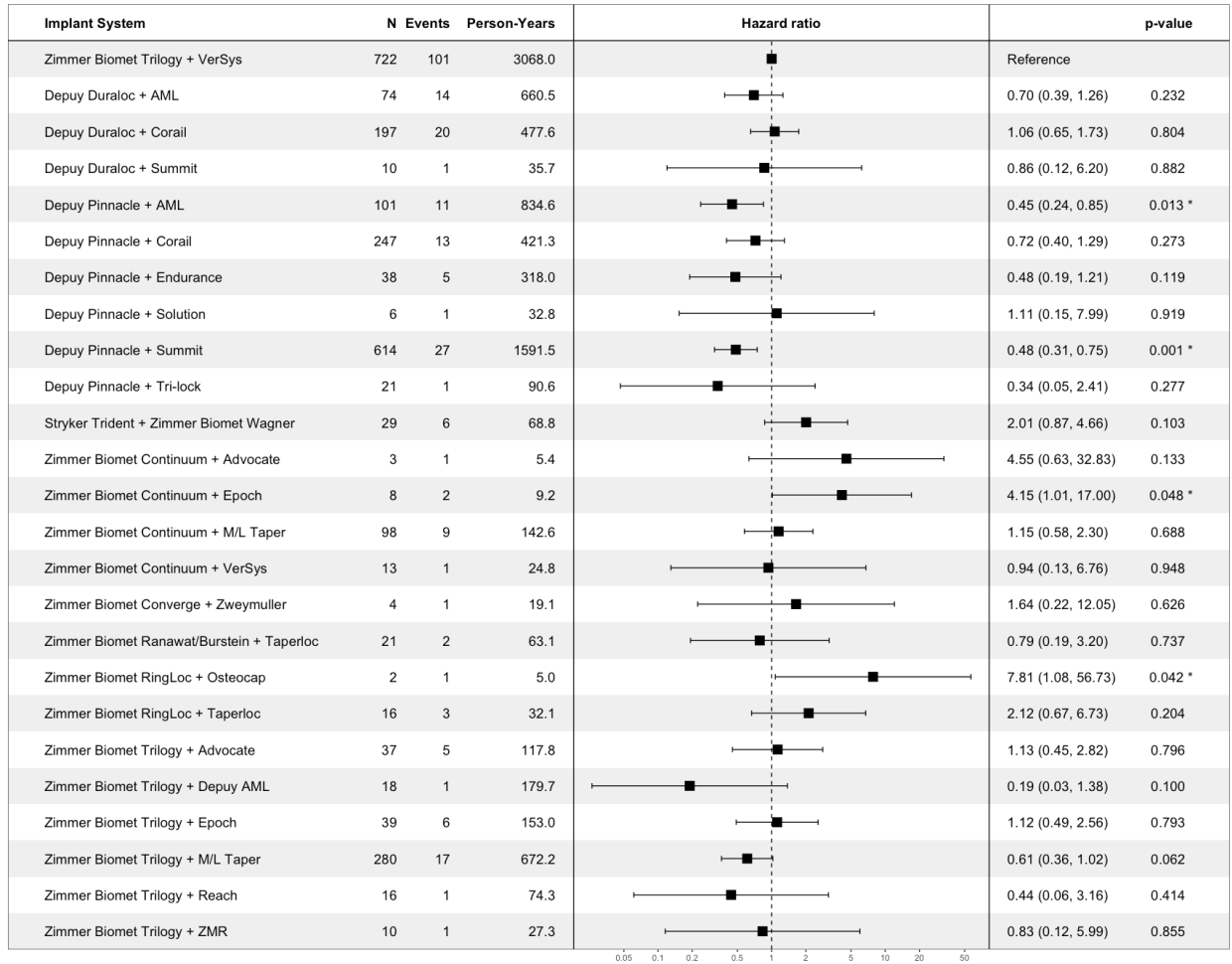

**Figure S4. Summary of Cox proportional hazards analysis of the risk of mechanical failure for each hip implant system.** The table on the left lists the number of patients implanted with each system, the number of mechanical failure events observed for each, and the total person-years of data available. The forest plot displays the corresponding hazard ratio, with the hazard ratio (95% confidence interval) and p-value listed in the table to the right. Note that this figure only shows implant systems for which at least one mechanical failure event was detected.

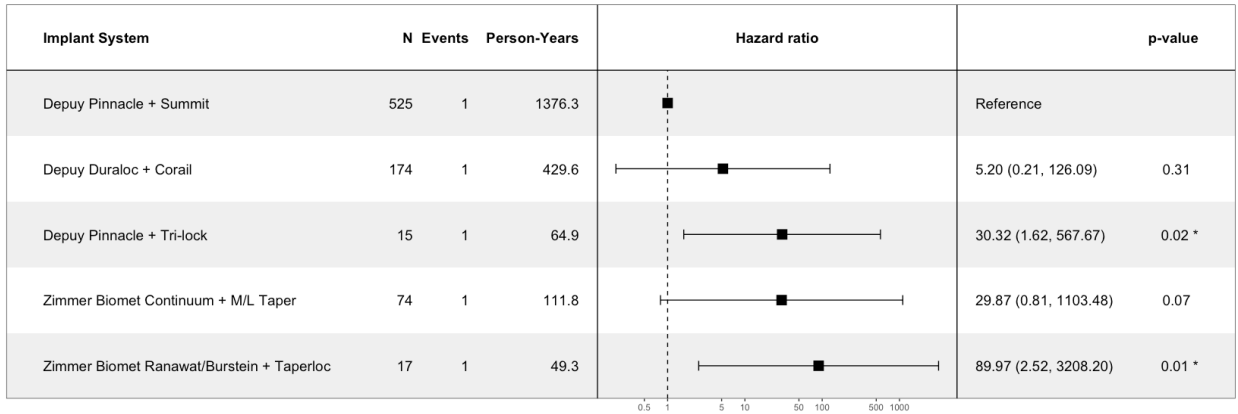

**Figure S5. Summary of Cox proportional hazards analysis of the risk of particle disease for each hip implant system.** The table on the left lists the number of patients implanted with each system, the number of particle disease events observed for each, and the total person-years of data available. The forest plot displays the corresponding hazard ratio, with the hazard ratio (95% confidence interval) and p-value listed in the table to the right. Note that this figure only shows implant systems for which at least one particle disease event was detected.

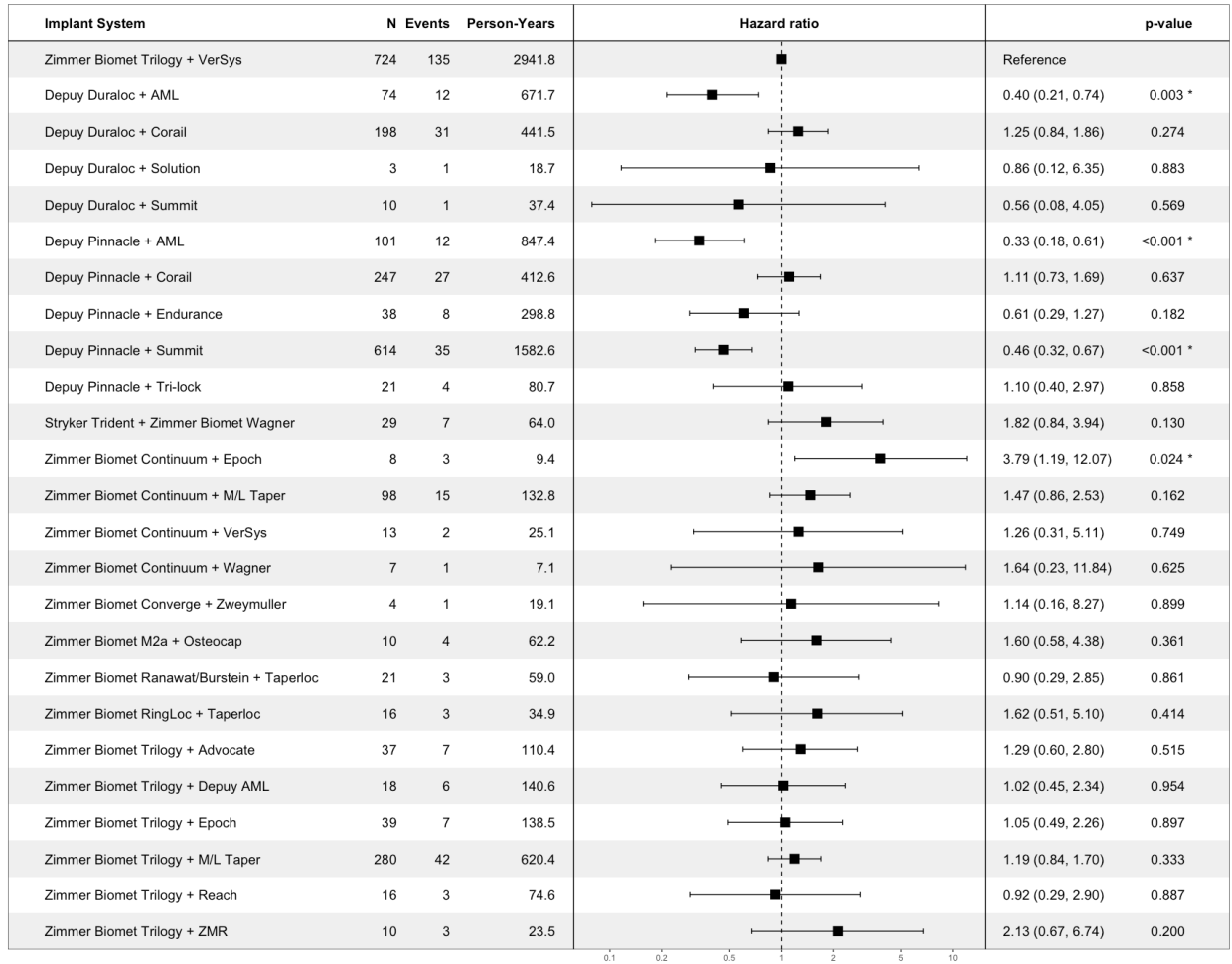

**Figure S6. Summary of Cox proportional hazards analysis of the risk of radiographic abnormality for each hip implant system.** The table on the left lists the number of patients implanted with each system, the number of radiographic abnormality events observed for each, and the total person-years of data available. The forest plot displays the corresponding hazard ratio, with the hazard ratio (95% confidence interval) and p-value listed in the table to the right. Note that this figure only shows implant systems for which at least one radiographic abnormality event was detected.

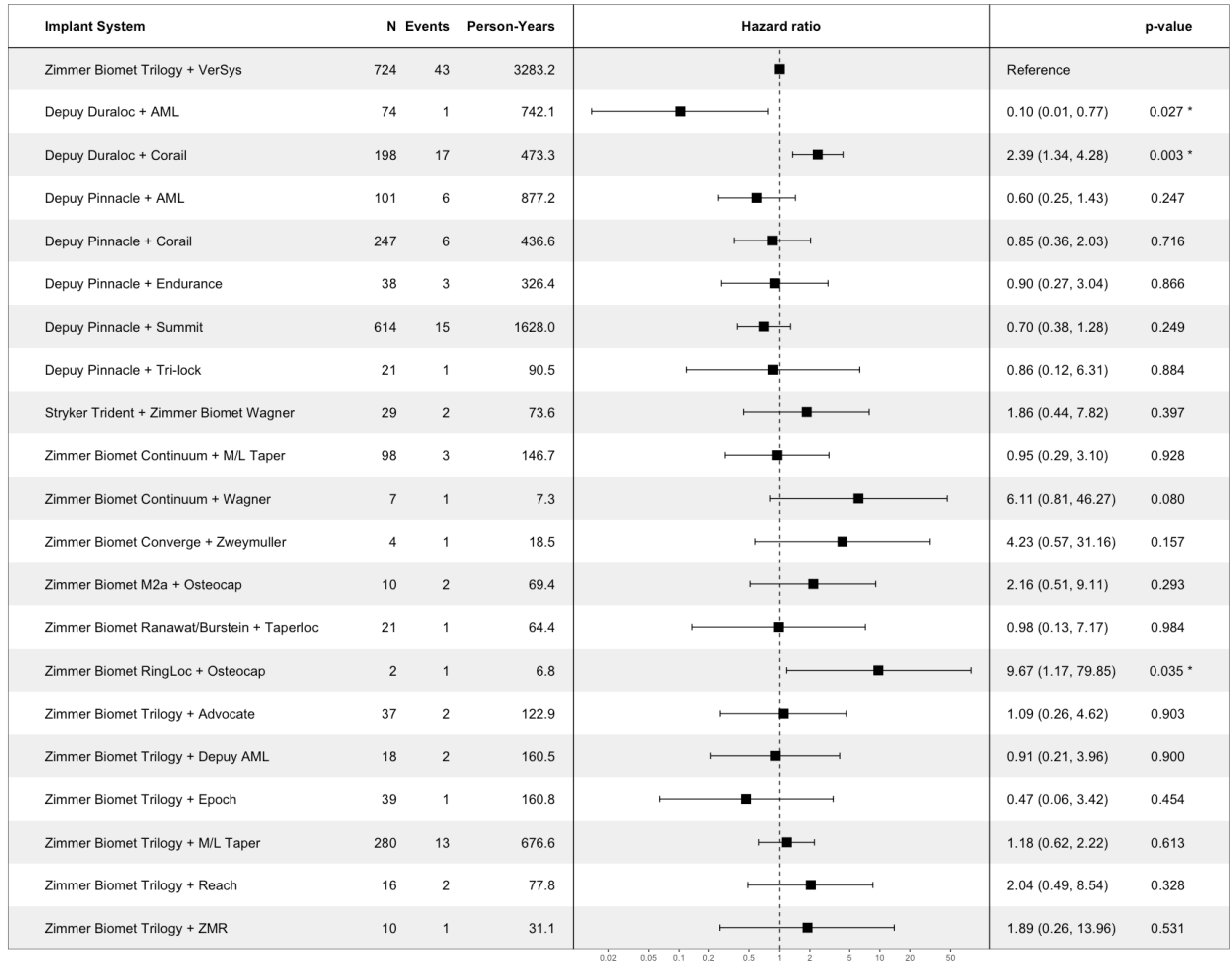

**Figure S7. Summary of Cox proportional hazards analysis of the risk of infection for each hip implant system.** The table on the left lists the number of patients implanted with each system, the number of infection events observed for each, and the total person-years of data available. The forest plot displays the corresponding hazard ratio, with the hazard ratio (95% confidence interval) and p-value listed in the table to the right. Note that this figure only shows implant systems for which at least one infection event was detected.

## Post-implant hip pain is associated with implant system

Table S1 lists the model coefficient estimate, incident rate ratio (IRR), lower and upper 95% confidence interval bounds, and p-values for the negative binomial model of hip pain in the year post-THA.

**Table S1.** Negative binomial model coefficients, IRR (95% confidence interval) and p-value for hip pain in the year after THA.

| Variable | Estimate | IRR (95% CI) | p-value |
|---|---|---|---|
| (Intercept) | 0.828 | 2.290 (1.455-3.604) | < 0.001 * |
| **Implant System** | | | |
| Zimmer Biomet Trilogy + VerSys | Reference | | |
| Depuy Duraloc + AML | -3.415 | 0.033 (0.012-0.091) | < 0.001 * |
| Depuy Duraloc + Corail | 0.238 | 1.268 (1.020 -1.577) | 0.033 * |
| Depuy Duraloc + Summit | -1.827 | 0.161 (0.040-0.652) | 0.011 * |
| Depuy M2A + Osteocap | 0.408 | 1.504 (0.640-3.537) | 0.350 |
| Depuy Pinnacle + AML | -2.334 | 0.097 (0.057-0.164) | < 0.001 * |
| Depuy Pinnacle + Corail | -0.521 | 0.594 (0.475-0.742) | < 0.001 * |
| Depuy Pinnacle + Endurance | -3.431 | 0.032 (0.008-0.137) | < 0.001 * |
| Depuy Pinnacle + Solution | -0.417 | 0.659 (0.199-2.187) | 0.496 |
| Depuy Pinnacle + Summit | -1.020 | 0.361 (0.301-0.432) | < 0.001 * |
| Depuy Pinnacle + TriLock | -0.524 | 0.592 (0.300-1.171) | 0.132 |
| Zimmer Biomet Continuum + Epoch | 0.530 | 1.700 (0.679-4.257) | 0.258 |
| Zimmer Biomet Continuum + M/L Taper | 0.723 | 2.061 (1.561-2.720) | < 0.001 * |
| Zimmer Biomet Continuum + VerSys | 0.589 | 1.802 (0.881-3.685) | 0.107 |
| Zimmer Biomet Continuum + Wagner | -0.352 | 0.703 (0.242-2.047) | 0.518 |
| Zimmer Biomet Trilogy + Depuy AML | -0.411 | 0.663 (0.321-1.367) | 0.266 |
| Zimmer Biomet Trilogy + Epoch | 0.313 | 1.368 (0.884-2.117) | 0.160 |
| Zimmer Biomet Trilogy + M/L Taper | 0.399 | 1.490 (1.234-1.799) | < 0.001 * |
| Zimmer Biomet Trilogy + Reach | -0.345 | 0.708 (0.335-1.499) | 0.367 |
| Zimmer Biomet Trilogy + Wagner | 0.607 | 1.834 (1.120-3.004) | 0.016 * |
| Zimmer Biomet Trilogy + ZMR | -0.237 | 0.789 (0.319-1.953) | 0.608 |
| Other system | -0.278 | 0.757 (0.504-1.138) | 0.181 |
| **Charlson Comorbidity Index** | | | |
| None | Reference | | |
| Low | 0.141 | 1.151 (0.956-1.386) | 0.137 |
| Moderate | 0.040 | 1.040 (0.809-1.338) | 0.758 |
| High | 0.234 | 1.264 (0.997-1.601) | 0.053 |
| **Age** | | | |
| 40-49 years | Reference | | |
| 50-59 years | -0.035 | 0.965 (0.794-1.174) | 0.723 |
| 60-69 years | 0.050 | 1.051 (0.870-1.269) | 0.606 |
| 70-79 years | -0.062 | 0.940 (0.768-1.151) | 0.549 |
| 80+ years | -0.219 | 0.803 (0.632-1.020) | 0.072 |
| **Sex** | | | |
| Female | Reference | | |
| Male | 0.010 | 1.010 (0.898-1.137) | 0.862 |
| **Race** | | | |
| Asian | Reference | | |
| Black | 0.108 | 1.114 (0.726-1.709) | 0.620 |
| Native American | -0.171 | 0.843 (0.171-4.162) | 0.834 |
| Other | 0.394 | 1.482 (1.047-2.099) | 0.027 * |
| Pacific Islander | 0.357 | 1.430 (0.598-3.420) | 0.422 |
| Unknown | 0.053 | 1.055 (0.684-1.626) | 0.810 |
| White | -0.030 | 0.970 (0.751-1.253) | 0.817 |
| **Ethnicity** | | | |
| Hispanic | Reference | | |
| Not Hispanic | 0.036 | 1.036 (0.746-1.439) | 0.832 |
| Unknown | -0.719 | 0.487 (0.328-0.723) | < 0.001 * |
| **Other covariates** | | | |
| Pain in year prior to THA | 0.071 | 1.073 (1.049-1.099) | < 0.001 * |
| Follow-up time | -0.001 | 0.999 (0.999-1.000) | 0.001 * |

## Relation extraction system performance

Table S2 details Pain-Anatomy extraction performance given 150 - 50,000 weakly labeled training documents. Here we see performance improvements up to +9.2 F1 points over soft majority vote as we increase the scale of weakly labeled data provided to the deep learning model.

**Table S2.** `Pain-Anatomy` Relation Extraction Performance

| Model | Training Set Size (Number of Documents) | | | | | | | | | | | |
|---|---|---|---|---|---|---|---|---|---|---|---|---|
| | 150 | | | 5K | | | 10K | | | 50K | | |
| | P | R | F1 | P | R | F1 | P | R | F1 | P | R | F1 |
| Supervised-LSTM ✦ | 72.5 | 78.4 | 75.4 | - | - | - | - | - | - | - | - | - |
| Soft Majority Vote | **81.4** | 64.8 | 72.2 | - | - | - | - | - | - | - | - | - |
| Weakly Supervised LSTM | 68.4 | 81.8 | 74.5 | 75.1 | 80.5 | 77.7 | 76.4 | 80.9 | 78.6 | 80.2 | **82.6** | **81.4** |

✦ Uses hand-labeled training data

Blue highlighting Highest achieved value for metric (P, R, F1)

Table S3 contains a non-exhaustive list of example terms for each Implant-Complication category. These terms form disjoint sets for each Implant-Complication sub-category.

**Table S3.** Example Terms for `Implant-Complication` Subcategories

| Subcategory | Terms |
| --- | --- |
| Mechanical failure | hardware loosening, crooked, asymmetrically seated |
| Revision | reoperation, removals, revision, rebuilt, hardware removal |
| Component wear | polyethylene wear, worn, wearing, bearing surface wear, debonding |
| Infection | infection, septic, abscess, re-infected |
| Particle disease | particle disease, metal ion toxicity, metallosis |
| Radiographic abnormality | lucencies, pedestals, heterotopic calcifications, spurs |